# Eurasian Economic Union: Current Concept and Prospects

**Larisa Kargina,**
*Professor, Russian University of Transport,*
*Moscow, Russia*

**Mattia Masolletti,**
*Associate Professor, NUST University*
*Rome, Italy*

**Abstract**

The authors of the article analyze the content of the Eurasian integration, from the initial initiative to the modern Eurasian Economic Union, paying attention to the factors that led to the transition from the Customs Union and the Single Economic Space to a stronger integration association. The main method of research is historical and legal analysis.

### 1. Introduction

The experience of Eurasian integration in the post-Soviet space, obtained through numerous initiatives, mistakes and unresolved issues, has become a necessary basis for the formation of a functional functioning association. In addition, it contributed to the identification of countries among the entire Eurasian space whose interests coincide in modern geopolitical and economic realities. The current stage of Eurasian integration is the formation of the Eurasian Economic Union on the basis of the Customs Union and a Single Economic Space.

### 2. Main part

The idea of creating such an association belongs to the former President of the Republic of Kazakhstan Nursultan Nazarbayev, who in 1994 had put forward the initiative of rapprochement of states in the format of the Eurasian Union. At that time, Kazakhstan, like other countries of the former USSR, had many representatives of various ethnic groups in its national composition. To strengthen the newly formed state, it was necessary to increase the welfare of the people and ensure equal rights of all nationalities in order to prevent the outflow of the population. At the same time, it was extremely important to establish and maintain friendly relations with the former Soviet republics, to establish cooperation in the field of security.

The peculiarity of the project proposed by Nazarbayev was to combine the national interests of Kazakhstan with the priority tasks of the development of other states of the Eurasian space. As a result, the Eurasian idea became one of the foundations of foreign policy and determined the vector of development of Kazakhstan. The President of Kazakhstan considered it necessary to achieve a high level of development of the Eurasian states by combining their economic (see *Fig.1*), transport and resource potentials.

In order to further develop integration processes in the post-Soviet space, the experience of international associations, primarily the European Union, was studied. The new format of the association, similar to the EU and unlike the Economic Union of 1993, implied the formation of supranational bodies, whose decisions are binding on the member states. However, it was emphasized that domestic and foreign policy issues should be within the competence of a sovereign state without delegation to the supranational level.

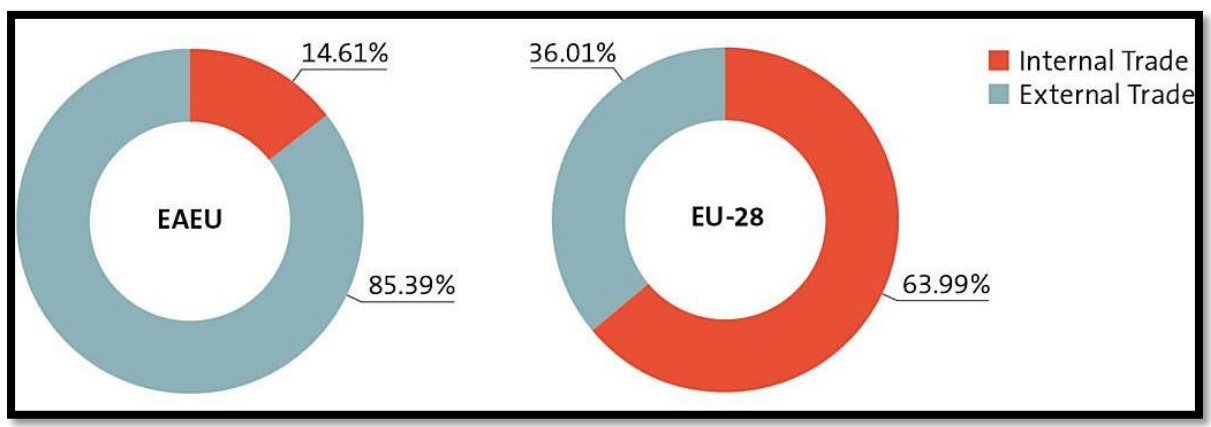

Fig.1. External and Internal Trade: the EU and EAEU in comparison (in 2017)

*Source: Eurostat 2017, Eurasian Economic Commission 2017* [6; 7]

At the first stage, Nazarbayev's initiative had assumed the formation of the core of integration, that is, the unification of the states, which are most ready for mutual rapprochement - Kazakhstan and Russia. This format of the union was supposed to become the center of attraction of the post-Soviet countries. Later, the main aspects of the construction of the Eurasian Union have been included into the book 'The Eurasian Union: ideas, Practice, prospects for 1994-1997'. In contrast to the previous concepts of eurasianism, which mainly consider it from a geographical, geopolitical or civilizational perspectives, Nazarbayev's ideas were of an applied, practical nature. At the same time, they have not just taken into account the current situation in the post-Soviet space at that time, but were also formulated for the first time taking into account the further development of globalization processes.

The principal feature of the project proposed by Nazarbayev was an appeal to the Eurasian idea, which implied the presence of an ideological component in his initiative. This meant that the economic integration of the states was justified not only by the need for economic cooperation, but also by the historical trends in the development of the entire Eurasian space, objectively aimed at finding optimal forms and mechanisms for preserving the ties that have existed between the Eurasian countries for centuries. Thus, the main provisions of the strategy for the creation of the Eurasian Union, much later set out by Nazarbayev in the newspaper 'Izvestia', they were as follows:

1) The Eurasian Union should initially be created as a competitive global economic association.

2) The Eurasian Union should be formed as a strong link linking the Euro-Atlantic and Asian areas of development.

3) The Eurasian Union should be formed as a self-sufficient regional association that will be part of the new global monetary and financial system.

4) The geo-economic, and in the future, the geopolitical maturation of the Eurasian integration should be exclusively evolutionary and voluntary.

5) The creation of the Eurasian Union is possible only on the basis of broad public support.

However, at that time, Nazarbayev's proposal was not practically implemented. In the early 1990s, the post-Soviet states, as already noted, faced serious economic problems, the solution of which depended on the future of these countries. In such conditions, integration processes did not seem to them to be an urgent issue, since first of all it was necessary to stabilize the economic situation. In addition, the independence of the States and the memory of strict centralized management did not allow the proposal to create supranational bodies to be perceived with interest. That is, the development of economic cooperation at that time was opposed to the strengthening of national independence and sovereignty. Despite the fact that Nazarbayev's initiative on the creation of the Eurasian Union in the 1990s had not found wide support, its role in the subsequent processes in the post-Soviet space was significant.

By the beginning of the 2000s, numerous attempts at rapprochement had identified a circle of countries ready to unite and carry out coordinated actions in various spheres of the economy. By this time, such states as Belarus, Kazakhstan and Russia have largely coped with the crises of the 1990s and have become the most economically developed, open and interested in restoring trade relations.

Nazarbayev's initiative was supported by Russia, namely by Vladimir Putin, who took office as President of the Russian Federation. Probably, the integration processes would not have developed with such intensity or remained theoretical inventions of traditional and modern Eurasians, if the new leader had not had a serious impact on the practical implementation of the idea of the Eurasian dialogue in the new historical period.

In the process of economic reforms aimed at improving the economy after the unfavorable 1990-s, developing the business climate and modernization, the need for close economic cooperation with neighboring countries was realized, on which not only the dynamics of foreign trade depended, which, of course, affected the welfare of the population, but also national security. It is obvious that in the XXI century, effective economic cooperation can only be carried out on a mutual basis, so Russia, like other countries participating in the rapprochement process, began difficult negotiations to coordinate interests and conditions of interaction and made concessions on some issues. Thus, the factor of political will played an important role in the formation of the modern concept and practice of Eurasian integration.

Assessing the political course of the Russian President Vladimir Putin as a whole, as well as actions aimed at updating the Eurasian issue, we can distinguish three bases on which the Eurasian practice is based:

- Civilizational and historical basis. In its historical retrospect, Russia is perceived as a single civilizational space that unites numerous ethnic groups, whose self-worth has been preserved for centuries thanks to the 'state-civilization' model, regardless of the historical period and its characteristic values and political structure. At the same time, the dominant principle-the continuity of history-is the optimal assessment of the entire historical path of Russia, since it allows, without giving preference to any political theory underlying a particular model of the state structure, to identify and apply the best practices of managing the Eurasian state.

Thus, in an interview with American journalist Charlie Rose for the CBS and PBS TV channels, Vladimir Putin expressed his desire to preserve the common humanitarian space in the region 'to make sure that state borders do not arise, so that people can freely communicate with each other, so that a joint economy develops, using the advantages that we inherited from the former Soviet Union' [3]. Such advantages include the presence of 'a common infrastructure, a single railway transport, a single road network, a single energy system' and 'the great Russian language, which unites all the former republics of the Soviet Union and gives us obvious competitive advantages in promoting various integration projects in the territory of the post-Soviet space' [3].

- Economic basis. Of course, economic pragmatism can be traced in the Eurasian practice, which today primarily explains the project of forming the EAEU. Over the past 15 years, the economic climate in Russia had been improved significantly (we will not consider the pandemic crisis separately). At the same time, it is obvious that entering the global market with domestic products requires high quality standards and competitive prices. To achieve these goals, technological and resource potentials are being developed within the framework of the Eurasian Economic Union, production costs are being reduced, and the position of producers of the member states in the single internal market is being strengthened. In the future, joint access to foreign markets is carried out, among other things. Thus, over time, the presence of domestic goods, services and capital in the global economy increases significantly.

- The geopolitical basis. Despite a significant share of economic pragmatism, the economy is not an end in itself. Through established cooperation relations with partners in the EAEU, improving the business climate, Russia's position in the CIS space is strengthened, security is ensured at the borders with neighboring states, and the influence and authority of the state in the world is increased. The priority of the development of Eurasian integration is fixed in the Concept of the Foreign Policy of the Russian Federation.

It should be noted that, unlike classical Eurasians, Vladimir Putin does not oppose Russia and, in general, the EAEU to the West. On the contrary, he sees the future of the Eurasian space in close connection with Europe. This is how he actualizes the thesis of Charles de Gaulle about the creation of 'Europe from the Atlantic to the Urals', that is in Putin's interpretation 'a common economic space from Lisbon to Vladivostok'. However, the elimination of trade and other restrictions between the two major economic centers can only be implemented on an equal and mutually beneficial basis.

Throughout the entire process of the Eurasian economic integration, the interest and consistent practical support of the Eurasian project by Belarus in the person of President Alexander Lukashenko can be traced. Thus, in his address to the heads of the member states of the Eurasian Economic Union, sent in connection with Belarus' chairmanship in the EAEU bodies in 2015, the desire to promote the four fundamental economic freedoms of the EAEU had been once again confirmed: the freedom of movement of goods, services, capital and labor, maximum liberalization of the conditions of economic activity within the EAEU, including through the complete abolition of exemptions and restrictions in the movement of goods.

In addition, it is Russia and Belarus that have formed the most advanced integration association in the post-Soviet space today, where not only economic but also social freedoms are ensured - the Union State of Belarus and Russia.

It is assumed that in the situation of a political conflict between the West and Russia, the support of Belarus, its consistent and clear position on the promotion of the Eurasian project will give a new impetus to economic integration, including taking into account the policy of import substitution.

So, having considered the main aspects of the development of the Eurasian idea, set forth by the founders of eurasianism, its transformation into a modern Eurasian concept, implemented already at the official level during the formation of the Eurasian Economic Union, taking into account the accompanying global geopolitical and geo-economic conditions, it is advisable to draw the following conclusions.

In the conditions of the collapse of the Russian Empire at the beginning of the XX century and the USSR at the end of the XX century, the rethinking of the role of the Eurasian states in various historical periods is quite a natural process aimed at finding the essence of the co-evolution of various peoples and cultures within the unique, geographically isolated Eurasian space and trying to predict or suggest a further vector of its development.

Classical Eurasians come to the following conclusion: in the Eurasian space, taking into account the peculiarities of its geographical location and resource potential, the existence of a single powerful political entity is historically conditioned. In different periods, such formations were the Empire of the Huns, the Turkic Khaganate, the Empire of Genghis Khan, the Russian Empire and the USSR. Despite the trend that has emerged in recent centuries of fragmentation of the common territory of the Eurasian space into various smaller state entities, geographically, historically and economically (bearing in mind the resource potential) this territory remains united, capable of self-sufficiency. A significant role in the integration potential is played by a common history, including the history of joint protection of the Eurasian space during various wars, a similar way of life, mentality and culture, as well as a single nation formed over the centuries, consisting of many ethnic groups and peoples, but united by the widely spread Russian language. Therefore, the creation of various kinds of unions and associations of states is currently not only natural, but also vital for the population of these countries experiencing the pressure of globalization processes. The formation of the EAEU in 2015 also corresponds to this concept, since it is aimed at improving the well-being of citizens of the member states by combining domestic markets, resources, establishing cooperative ties and conducting a coordinated macroeconomic policy.

The analysis of classical and modern eurasianism demonstrates the conditionality of the differences associated with modern geopolitical and geo-economic realities.

Thus, if the founders of the Eurasian idea primarily opposed it to the Western vector of development and, in general, to the West, at present, not only the possibility, but also the need for the coexistence of integration processes in Europe and Eurasia, including through the creation of a common economic space, is being asserted. This thesis has been repeatedly voiced in the political discourse of official representatives of the EU and the EAEU member states.

However, in the current difficult situation of the exchange of sanctions between Russia and the West, the mutual lifting of restrictions in the implementation of economic and other activities is very unlikely. At the same time, the need to liberalize the trade regime between them becomes more obvious. After all, in the current historical period, some countries of the post-Soviet space are forced to choose between European and Eurasian integration. And if, for example, the Baltic countries unambiguously integrated into the EU, and Armenia joined the Eurasian Economic Union in 2015, and relatively painlessly, then the lack of an unambiguous position in Ukraine on this issue led to a civil war. Thus, at present, European and Eurasian integration are rather opposed to each other, therefore, this distinction between classical and modern eurasianism is currently conditional.

In addition, a distinctive feature of the modern Eurasian concept is its strict focus on economic integration. If the founders of the Eurasian idea talked more about cultural pluralism, the historical predestination of the cohabitation of different ethnic groups on the same territory, then at the end of the XX – beginning of the XXI century, the self-worth of each people is already fixed at the legislative level, and due to the recent ambiguous imperial and Soviet past, discussions of political or social integration are very sensitive and therefore are practically not brought to the official level.

However, it seems that the Eurasian integration is a more complex and complex process. The creation of the Eurasian Economic Union is only the basis for building a broader Eurasian strategy, on the basis of which, if successfully implemented, there will be a deepening of integration in the military, political, social, migration and humanitarian spheres. In this case, in the long term, the consequences of such a rapprochement are likely to include:

- the formation of the Eurasian identity as the self-identification of the peoples of the Eurasian space, united by a common economic system and culture, value orientations and a certain system of protection of these values, that is, an integrated security system;

- the development of the Eurasian mentality as a stable set of attitudes and predispositions to perceive the world in a similar way, possibly different from the worldview of residents of other global regions.

### 3.    Conclusion

It should be concluded that the modern processes of Eurasian integration are taking place in line with the classical idea of eurasianism, which proves its relevance. Today, as a hundred years ago, the regularity of the joint development of the peoples of the Eurasian states in the common space is confirmed. Therefore, in the conditions of the formation of a polycentric world order, the struggle for influence in the centers of power between the existing global leaders, these countries need to become an independent and equal subject of international relations in order to preserve their sovereignty, political independence, the heritage of history, science and culture, modernization and development of innovations in the economy, which is possible only by combining their potentials and peaceful coexistence, complementarity and interpenetration of their peoples.

**References**

[1] Nazarbayev N.A. (1997) The Eurasian Union: ideas, practice, prospects. 1994-1997. Moscow: Foundation for the Promotion of Social and Political Sciences.

[2] Address of the President of the Republic of Belarus Alexander Lukashenko to the heads of the member States of the Eurasian Economic Union dated January 1, 2015 // Official website of the Eurasian Economic Commission. - URL: http://www.eurasiancommission.org/ru/nae/news/Pages/21-01-2015.aspx

[3] Vladimir Putin's interview with American journalist Charlie Rose for CBS and PBS TV channels from September 29, 2015. URL: http://kremlin.ru/events/president/news/50380

[4] Putin V. (2011) A new integration project for Eurasia — the future that is being born today // The newspaper 'Izvestia', dated November 3, 2011.

[5] Nazarbayev N.A. (2011) The Eurasian Union: From an idea to the history of the future // The newspaper 'Izvestia'.

[6] Eurostat (2017). External and Internal Trade in the EU.

[7] Eurasian Economic Commission (2017). External and Internal Trade in the EAEU.